\documentclass[showpacs,showkeys, aps, prb, twocolumn, superscriptaddress]{revtex4-1}

\usepackage[sort&compress]{natbib}
\usepackage{graphicx}
\usepackage{dcolumn}
\usepackage{bm}
\usepackage{amsmath,empheq}
\usepackage{setspace}
\usepackage{hyperref}
\hypersetup{colorlinks=true, linkcolor=blue, citecolor=cyan}

\begin{document}

\title{Long range magnetic order in hydroxide layer doped (Li$_{1-x-y}$Fe$_{x}$Mn$_{y}$OD)FeSe} 

\author{Brandon Wilfong}
\affiliation{Department of Chemistry and Biochemistry, University of Maryland, College Park, Maryland 20742, United States}
\affiliation{Quantum Materials Center, University of Maryland, College Park, Maryland 20742, United States}
\author{Xiuquan Zhou}
\affiliation{Department of Chemistry and Biochemistry, University of Maryland, College Park, Maryland 20742, United States}
\affiliation{Materials Science Division, Argonne National Laboratory, Lemont, Illinois 60439, United States}
\author{Huafei Zheng}
\affiliation{Department of Chemistry and Biochemistry, University of Maryland, College Park, Maryland 20742, United States}
\author{Navneeth Babra}
\affiliation{Department of Chemistry and Biochemistry, University of Maryland, College Park, Maryland 20742, United States}
\author{Craig M. Brown}
\affiliation{NIST Center for Neutron Research, National Institute of Standards and Technology, Gaithersburg, MD 20899}
\author{Jeffrey W. Lynn}
\affiliation{NIST Center for Neutron Research, National Institute of Standards and Technology, Gaithersburg, MD 20899}
\author{Keith M. Taddei}
\affiliation{Diffraction Group, Neutron Sciences Division, Oak Ridge National Laboratory, Oak Ridge, Tennessee 37831, United States}
\author{Johnpierre Paglione}
\affiliation{Quantum Materials Center, University of Maryland, College Park, Maryland 20742, United States}
\affiliation{Department of Physics, University of Maryland, College Park, Maryland 20742, United States}
\author{Efrain E. Rodriguez}
\email{efrain@umd.edu}
\affiliation{Department of Chemistry and Biochemistry, University of Maryland, College Park, Maryland 20742, United States}
\affiliation{Quantum Materials Center, University of Maryland, College Park, Maryland 20742, United States}


\begin{abstract}
The (Li$_{1-x}$Fe$_{x}$OH)FeSe superconductor has been suspected to exhibit long-range magnetic ordering due to Fe substitution in the LiOH layer. However, no direct observation such as magnetic reflection from neutron diffraction has be reported. Here, we use a chemical design strategy to manipulate the doping level of transition metals in the LiOH layer to tune the magnetic properties of the (Li$_{1-x-y}$Fe$_{x}$Mn$_{y}$OD)FeSe system. We find Mn doping exclusively replaces Li in the hydroxide layer resulting in enhanced magnetization in the (Li$_{0.876}$Fe$_{0.062}$Mn$_{0.062}$OD)FeSe superconductor without significantly altering the superconducting behavior as resolved by magnetic susceptibility and electrical/thermal transport measurements. As a result, long-range magnetic ordering was observed below 12 K with neutron diffraction measurements. This work has implications for the design of magnetic superconductors for the fundamental understanding of superconductivity and magnetism in the iron chalcogenide system as well as exploitation as functional materials for next generation devices.
\end{abstract}

\pacs{}
\keywords{}
\maketitle


\section{Introduction}

The binary FeSe, with a superconducting critical temperature ($T_{c}$) of 8 K \cite{Hsu}, provides an excellent template to study exotic physical phenomena in iron-based superconductors due to its simple structure, ease of chemical manipulation and relatively high superconducting critical temperature. Amazingly, the $T_c$ of FeSe can be improved significantly to 42-46 K from 8 K after intercalation \cite{Burrard,Chen_LiOH, Sun_LiOH}, 37 K with applied pressure \cite{marga} or 65 K in the monolayer limit \cite{he}. The well-studied intercalated compound (Li$_{1-x}$Fe$_{x}$OH)FeSe ($T_c$ = 42-44 K) consists of a tetragonal layer of partially charged Li$_{1-x}$Fe$_{x}$OH ($x \approx$ 0.1-0.2) between the FeSe layers. Such a structure is stabilized by the partial charge transfer due to Fe doping on the Li site as well as hydrogen bonding from the LiOH layer to the Se atoms in the FeSe layers \cite{Chen_LiOH, Wang_HB, Zhang_LiOH}. The Fe substitution in the insulating hydroxide layer not only plays a crucial role in the enhancement of $T_c$, but also can induce exotic physical phenomena such as coexistence of magnetic order and superconductivity \cite{Lu_LiOH, Pachmayr, Lynn_LiOH}. Although many reports have shown signatures of magnetic order in this compound, there are no definitive signs to understand the true nature of its magnetism. The magnetic transition is intrinsic to the system with different reports attributing this transition to ferromagnetic ordering, canted antiferromagnetic ordering, and spin glassiness \cite{Lu_LiOH, Pachmayr, Topping_LiOH, Lynn_LiOH, Wu_LiOH}.  Despite their differences, all of these studies point toward the hydroxide layer as the source of the magnetic ordering. 

\begin{figure}[htb!]
	\includegraphics[width=0.95\columnwidth]{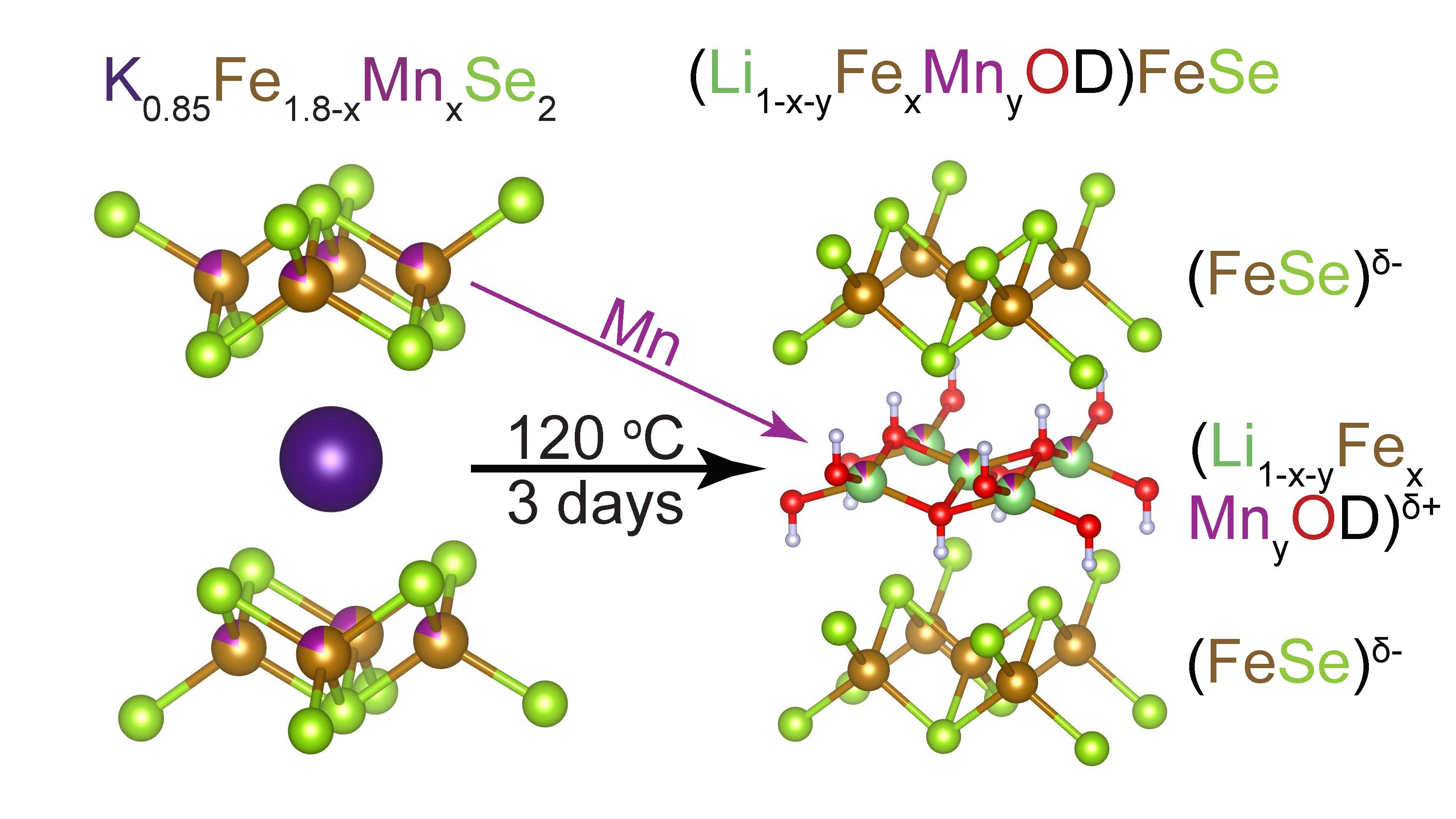}
	\caption{Synthetic scheme and results of targeted (Li$_{1-x-y}$Fe$_{x}$Mn$_{y}$OD)FeSe by converting Mn-doped KFe$_{1.8-z}$Mn$_{z}$Se$_2$ hydrothermally.}
	\label{fig:syn}
\end{figure}

In contrast to the uncertainties in (Li$_{1-x}$Fe$_{x}$OH)FeSe, there is a history of the observation of the coexistence of magnetism and superconductivity, mostly in rare-earth element containing compounds. The first examples of long-range magnetic order coexisting with superconductivity was observed in the ternary Chevrel phases RMo$_{6}$S$_{8}$ and RRh$_{4}$B$_{4}$ (R = lanthanide) where magnetic order arose from the lanthanide ion sublattice which was isolated from the superconducting sublattice. In these systems, both antiferromagnetism and ferromagnetism was observed coexisting with superconductivity depending on the lanthanide chosen for the magnetic sublattice \cite{lynn2, lynn3, moncton, lynn4}. The stabilization of magnetic order in these compounds was explained by dipolar electromagnetic interactions as the magnetic transition temperatures were below 1 K. Akin to these compounds were the later discovered borocarbides RNi$_{2}$B$_{2}$C where the magnetic ion sublattice now exhibits significant R-R exchange interactions pushing the magnetic phase transition much higher than in the Chevrel phases \cite{lynn5, gupta, lynn6}. The discovery of superconductivity in the cuprate family offered a new avenue for exploration. In these compounds, magnetic ion sublattices isolated from the CuO planes show low ordering temperature similar to the previously mentioned systems which can coexist with superconductivity. However, interestingly in these compounds, Cu ions (S = 1/2) of the CuO planes exhibited antiferromagnetic insulator behavior, but this antiferromagnetic order could be suppressed to induce superconductivity upon doping. \cite{paglione, lynn7, tokunaga, lynn9} The iron pnictide superconducting systems exhibit similar phenomena to the cuprate systems whereby magnetic order can coexist with superconductivity through an isolated magnetic ion sublattice and superconductivity arises from suppressing magnetic order of the iron pnictide layers through doping \cite{paglione, lynn8, xiao2009, Nandi2014, Bao2018, Stolyarov2018}. Unlike the related iron pnictide phases, FeSe exhibits no parent magnetic phase; \cite{Hsu, Bohmer} recent work has shown a high temperature nematic phase that precedes the superconducting phase \cite{Watson_nematic, Watson_nematic2, Chubukov_nematic}. Although no parent magnetically ordered phase exists, strong magnetic fluctuations have been observed in a wide range of temperatures in FeSe through neutron and NMR spectroscopy experiments \cite{Wang_magnetic, Bohmer, Rahn_magnetic, Wang_magnetic2}. Thus, the introduction of a magnetic spacer layer between FeSe may help to further reveal the role magnetic fluctuations play in superconductivity in the Fe-chalcogenide systems.

In addition to its interesting magnetism, experimental evidence has shown that (Li$_{1-x}$Fe$_{x}$OH)FeSe exhibits a Majorana Zero Mode (MZM) which plays a critical role in topological quantum computing applications \cite{Qin_MZM, Liu_MZM}.  Therefore, (Li$_{1-x}$Fe$_{x}$OH)FeSe is a perfect system to study and understand the coexistence of physical phenomena for possible applications in functional materials as well as quantum computing. 

Since all the aforementioned exotic phenomena emerge from the interactions between the superconducting FeSe$^{\delta -}$ and the insulating (Li$_{1-x}$Fe$_{x}$OH)$^{\delta +}$ layers, one may suggest to modify the hydroxide layer to induce new properties. Unfortunately, traditional solid-state reactions and methods will be insufficient to modify the solid solutions as these phases are metastable. Thus, we have developed a two-stage ion-exchange scheme to obtain (Li$_{1-x-y}$Fe$_{x}$Mn$_{y}$OD)FeSe by converting Mn-doped K$_{0.85}$Fe$_{1.8-z}$Mn$_{z}$Se$_2$ hydrothermally.  We manipulate the chemistry of the hydroxide layer in (Li$_{1-x}$Fe$_{x}$OD)FeSe through manganese doping to tune the magnetic properties without significantly altering the superconductivity in the system (Fig. \ref{fig:syn}). The addition of manganese to supplant Li and Fe in the tetrahedral hydroxide layer increases the effective spin and therefore the effective moment of layer to observe long range magnetic order.

\section{Methods}

\subsection*{Synthesis}
The synthesis of (Li$_{1-x-y}$Fe$_{x}$Mn$_{y}$OD)FeSe single crystals was perfomed via a two-step ion-exchange route similar to previous works \cite{Dong_LiOH, ma_LiOH, Zhou_LiOH}. High purity metallic K (Alfa Aesar, 99.5\%),  Mn granules (Alfa Aesar, 99.98\%), Fe granules (Alfa Aesar, 99.98\%), and Se shots (Alfa Aesar, 99.999\%) were used as raw materials. 

\par In order to incorporate transition metal doping, single crystals of K$_{0.85}$Fe$_{1.8-z}$Mn$_{z}$Se$_{2}$ were synthesized through a self-flux route with elemental mixture of K:Fe:Mn:Se in two nominal ratios of 0.85:(1.62/1.44):(0.18/0.36):2 were mixed in an argon glovebox sealed under vacuum in a double quartz ampoule. The quartz tubes were slowly heated at 50 $^{\circ}$C/hr to 1050 $^{\circ}$C, held at 1050 $^{\circ}$C to ensure a congruent melt, slowly cooled down to 550 $^{\circ}$C at a rate of 5 $^{\circ}$C/hr, and ended with natural cooling to room temperature. This method routinely produced 3 x 3 mm$^{2}$  plate-like single crystals. The hydrothermal ion-exchange were performed in 100 mL stainless steel autoclaves lined with Teflon cups. For each batch, select single crystals of K$_{0.85}$Fe$_{1.8-z}$Mn$_{z}$Se$_{2}$ with average total mass of approximately 2 grams, 1.2 mmol of selenourea (Sigma Aldritch, 98\%), 13 mmols of iron granules (Alfa Aesar, 99.98\%), 1.2 mmol of tin granules (used to regulate pH), 0.31 moles LiOH (anhydrous, Alfa Aesar, 98\%), and 40 mL of D$_{2}$O (Cambridge Isotope, 99.9\%) were loaded into the autoclave and purged under argon flow for 2 minutes before being tightly sealed. Each autoclave was heated to 120 $^{\circ}$C and held for 72 hours in a convection oven. Silver plate-like single crystals were recovered by washing away excess powders with D$_{2}$O, excess iron granules were easily recovered with a magnetic bar. Crystals were dried under vacuum overnight and stored in an Ar filled glovebox.

\subsection*{Magnetic and transport measurements}
All magnetic property measurements were carried out using a Quantum Design Magnetic Susceptibility Measurement System (MPMS-3) on powders and single crystals of (Li$_{1-x-y}$Fe$_{x}$M$_{y}$OD)FeSe samples. Zero-field-cooled (ZFC) and field-cooled (FC) measurements were taken from 1.8 to 300 K with various applied direct current (DC) magnetic fields. Isothermal magnetization measurements were taken from H = $\pm$ 7 T at numerous temperatures to probe the magnetic and superconducting state. Electrical transport measurements were performed on a 9 T Quantum Design Physical Property Measurement System (PPMS-9T) with temperatures from 1.8 to 300 K and fields up to 9 T  using a four-probe technique with current applied across the $ab$-plane due to the lamellar nature of the single crystals. Heat transport measurements were performed on a 14 T Quantum Design Physical Property Measurement System (PPMS-14T) over the range of 1.8 to 60 K using the relaxation technique \cite{QuantumDesign, HwangHC, BachmannHC}.

\subsection*{X-ray diffraction measurements}
Laboratory powder x-ray diffraction (PXRD) was collected using a Bruker D8 X-ray diffractometer with Cu K$\alpha$ radiation ($\lambda = 1.5406$ \AA, step size = 0.020$^{\circ}$, 2$\theta$ = 5 - 90$^{\circ}$) for phase identification. In order to aid in structural refinements associated with the (Li$_{1-x-y}$Fe$_{x}$Mn$_{y}$OD)FeSe system, as three different elements occupy the same crystallographic site, high resolution synchrotron X-ray diffraction was performed on powders of ground single crystals at Beamline 11-BM at the Advanced Photon Source at Argonne National Lab. Ground powders of single crystals were packed in 0.4 mm Kapton capillary tubes and sealed with epoxy. Diffraction data was collected between 0.5° and 46° with a step size of 0.0001° using a constant wavelength λ = 0.413964 \AA (30 keV) at 100 K. Rietveld refinements were performed using the TOPAS software suite \cite{TOPAS}.

\subsection*{Neutron diffraction measurements}
Neutron powder diffraction (NPD) data was collected at the NIST Center for Neutron Research (NCNR) BT-1 High Resolution Powder Diffractometer and Oak Ridge National Lab High Flux Isotope Reactor (HFIR) HB-2A \cite{calder}. Powder samples of ground single crystals of (Li$_{1-x-y}$Fe$_{x}$Mn$_{y}$OD)FeSe  were loaded into vanadium cans under helium exchange gas and loaded into a closed-cycle refrigerator (BT-1) or three-sample changer in 70mm Orange Cryostat (HB-2A). Low temperature diffraction data was collected at 2 K and 9 K and high temperatures at 50 K, for direct comparison in attempts to find magnetic satellite reflections, with Cu(311) ($\lambda = 1.54$ \AA) at BT-1 and Ge(115) ($\lambda = 1.54$ \AA) at HB-2A. Rietveld refinements were performed using the TOPAS software suite \cite{TOPAS}. In order to search for weak magnetic reflection, high intensity but coarse resolution diffraction measurements were performed on single crystals and ground single crystal powder on the BT-7 spectrometer at the NCNR using the position sensitive detector (PSD) with PG (002) ($\lambda = 2.359$ \AA) in a range of temperature from 2 - 50 K to search for magnetic transitions inside the superconducting regime.


\section{Evidence for long range magnetic order}

\begin{figure*}[t!]
	\includegraphics[width=1.75\columnwidth]{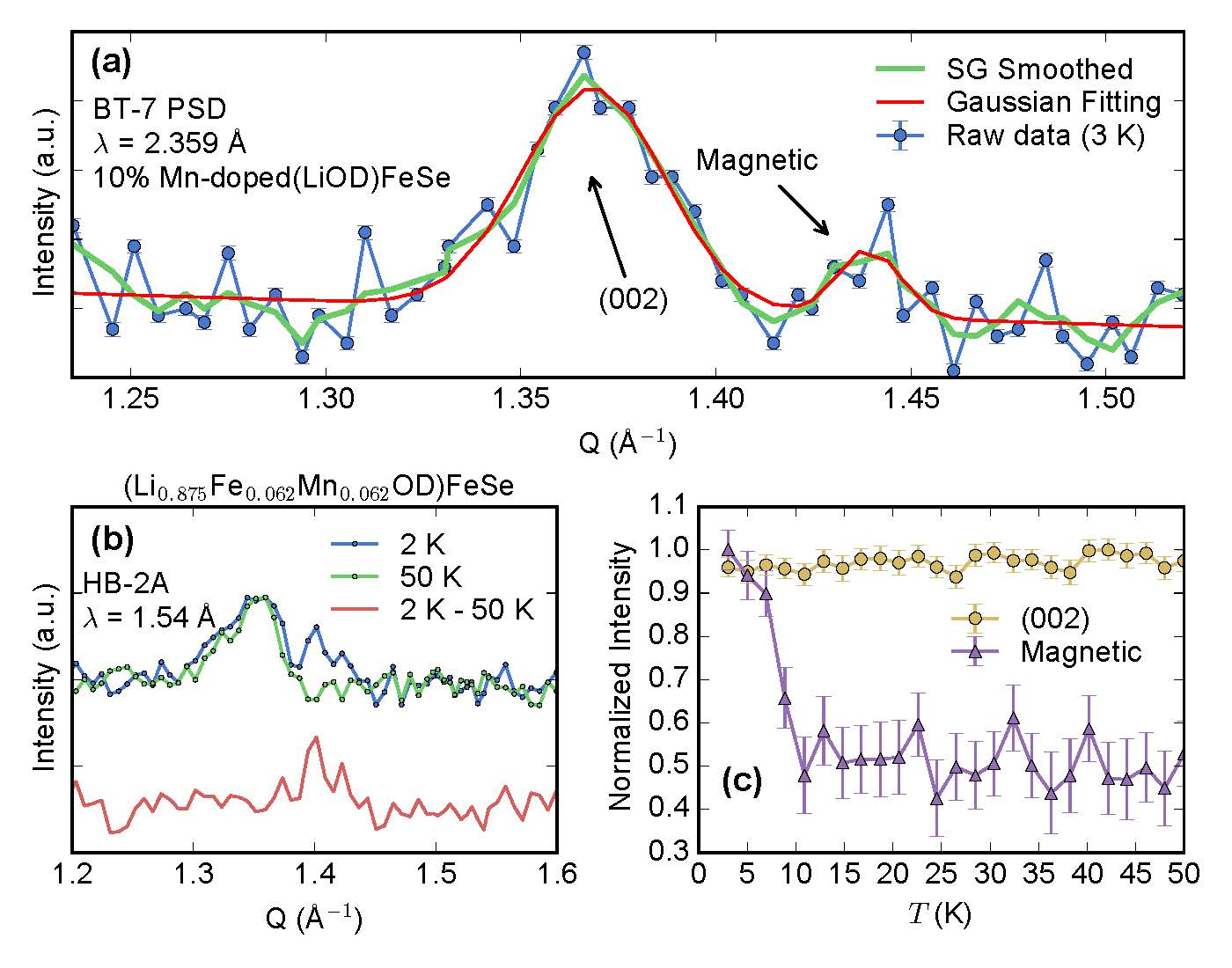}
 	\caption{Powder neutron diffraction data collected at BT-7 and HB-2A. a) Powder neutron diffraction data collected at 3 K at BT-7 using the PSD on a nominally 10\% Mn doped LiODFeSe sample of the (002) reflection showing a clear well-resolved magnetic peak proximate to the (002) reflection. This data smoothed using a Savitzsky-Golay filter and subsequently fit using a two-Gaussian model to extract temperature dependence. b) Powder neutron diffraction data on (Li$_{0.875(2)}$Fe$_{0.062(3)}$Mn$_{0.062(3)}$OD)FeSe at HB-2A showing the (002) reflection and magnetic peak at 2 K which is absent in the 50 K data. c) Temperature dependence on the integrated intensity of the two-Gaussian model on the data presented in a) showing a sharp decrease in the magnetic peak integrated intensity up to 10 K and temperature independence above 10 K. Uncertainties represent one standard deviation.}
  	\label{fig:neutron}
\end{figure*}

To date, all previous powder neutron diffraction data on (LiOD)FeSe samples made via the bottom-up synthesis and ion-exchange method do not exhibit observable long-range magnetic order \cite{Lu_LiOH, Pachmayr, Sun_LiOH, Zhou_LiOH, davies}. Small angle neutron scattering did reveal a characteristic ferromagnetic scattering below 12 K with a clear vortex scattering peak observed at $Q = 0.0077 \text{ }\mathrm{\AA}^{-1}$ under a 0.4 T applied field \cite{Lynn_LiOH}. Two more recent works on single crystals of the (LiOD)FeSe system via the ion-exchange method used inelastic neutron spectroscopy to investigate spin excitations below the superconducting temperature to reveal their importance in driving high critical temperatures in these compounds. \cite{ma_LiOH, pan_LiOH}

We have performed neutron powder diffraction on Mn-doped (LiOD)FeSe to observe long-range magnetic order. The results are shown in Figure \ref{fig:neutron} which span the use of two diffraction instruments and two different compositions of the Mn-doped (LiOD)FeSe system. To start, Figure \ref{fig:neutron}b shows the low $Q$ data accessible from the powder neutron diffraction data collected at HB-2A on Mn-doped (LiOD)FeSe and presented in Figure \ref{fig:xrd_npd}b/c in full. At $Q = 1.41 \text{ }\mathrm{\AA}^{-1}$, $d = 4.45 \text{ }\mathrm{\AA}$, a pronounced peak is observed at 2 K which is absent at 50 K; this peak is proximate to the nuclear (002) reflection but it was well resolved. This is the first time a satellite reflection has been observed in neutron diffraction for the (LiOD)FeSe system indicating the existence of some form of long-range order. We attribute this to the larger magnetic moment of Mn$^{2+}$ ($S$ = 5/2) compared to Fe$^{2+}$ (S = 2). As the magnetic peak appears distinct from the nuclear peaks, we can conclude the ordering in the system must be of the antiferromagnetic or ferrimagnetic type. Unfortunately, we observe only one magnetic peak in the data, most likely due to the small magnetic moment of the system or the small magnetic form factor in this compound. The full magnetic structure cannot be solved currently. 

In order to further understand how this magnetic peak behaved as a function of temperature, additional powder neutron diffraction data was collected on ground single crystals of nominally 10\% Mn-doped (LiOD)FeSe samples at BT-7 using the PSD. For these measurements, the (002) reflection was identified and measured as a function of temperature with a $\pm 0.2 \text{ }\mathrm{\AA}^{-1}$ window of collection using the PSD. Thus, the satellite reflection proximate to the (002) reflection observed in HB-2A data could be recorded concurrently. The raw data collected using the PSD at 3 K is shown in Figure \ref{fig:neutron}a displayed with a two-Gaussian model fit to the smoothed data using a third-order Savitzsky-Golay filter due to the coarse resolution of the raw data. The raw data and two-Gaussian fit show two well resolved peaks, one for the (002) reflection at $ Q = 1.37 \text{ }\mathrm{\AA}^{-1}$ and one for the magnetic peak at $ Q = 1.44 \text{ }\mathrm{\AA}^{-1}$, in close agreement to the HB-2A data. The slight discrepancy is likely due to crystallographic differences in the compounds. Figure \ref{fig:neutron}c shows the temperature evolution of the normalized integrated intensities calculated by the two-Gaussian model corresponding to the (002) reflection and magnetic peak. It is observed that the (002) reflection is temperature independent while the magnetic peak shows a sharp decrease in integrated intensity above 10 K in close agreement to the ordering temperature observed in magnetization and heat capacity data. We can conclude that the long-range magnetic ordering observed is intrinsic to the system with an ordering temperature around 9 K and has been observed for the first time by targeted design of the hydroxide layer in (LiOD)FeSe.

Futhermore, in the Mn-doped (LiOD)FeSe we have observed a high superconducting critical temperature paired with a relatively high magnetic transition temperature compared to rare-earth containing phases. This arises through similar means to the previously mentioned Chevrel, borocarbide, copper and iron-based systems whereby the magnetic sublattice and superconducting sublattice are isolated from another. However, we have done so without the need for rare-earth ions. Interestingly, in this system the addition of transition metal doping in the hydroxide layer both induces magnetic order as well as charge dopes the FeSe layers significantly raising the critical temperature of the compound. Thus, transition metal doped (LiOH)FeSe offers a tremendous platform for exploration of the role magnetism plays in stabilizing high temperature superconducitivity in the iron chalcogenides and for potential use in next generation devices.


\section{Hydrothermal synthesis and crystallographic results}

\begin{figure*}[t!]
	\includegraphics[width=1.75\columnwidth]{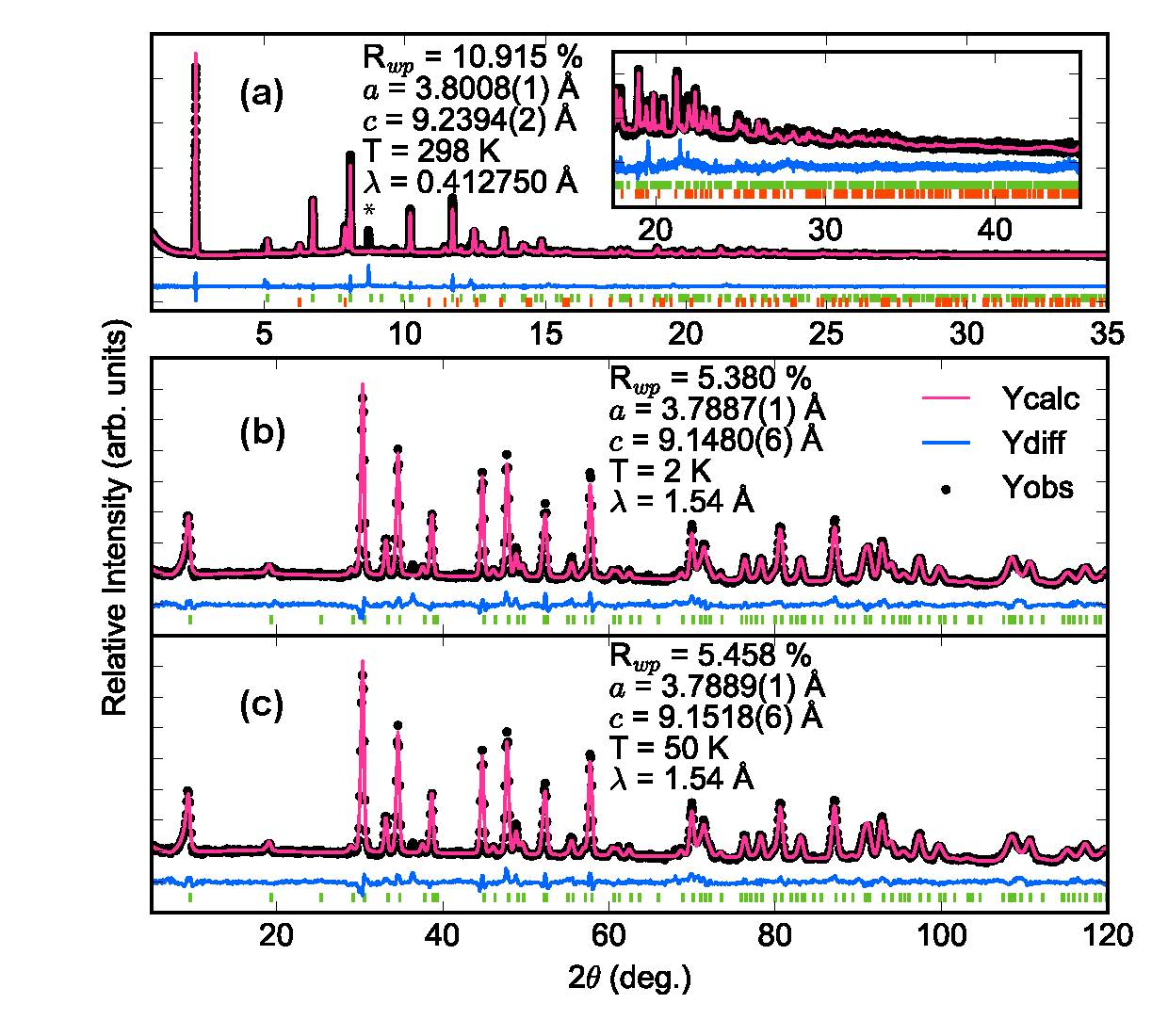}
	\caption{ a) High resolution synchrotron PXRD patterns for (Li$_{1-x-y}$Fe$_{x}$Mn$_{y}$OD)FeSe collected at room temperature, b) and c) show NPD data for (Li$_{1-x-y}$Fe$_{x}$Mn$_{y}$OD)FeSe at 2 K and 50 K respectively. Green tick marks represent the targeted tetragonal layered phase and orange tick marks represent impurity selenium. These are shown below the calculated, observed, and differences curves from Rietveld analysis. The asterisk represent a single observable impurity peak that does not match any observable peak in the NPD data.}
	\label{fig:xrd_npd}
\end{figure*}

As shown in Fig.\ref{fig:syn}, we first prepare precursors of K$_{0.85}$Fe$_{1.8-z}$Mn$_{z}$Se$_2$ with 10\% and 20\% nominal Mn doping level using direct elemental reactions from high-temperature. Our XRD analysis of K$_{0.85}$Fe$_{1.8-z}$Mn$_{z}$Se$_2$ show significant different lattice constants compared to K$_{0.85}$Fe$_{1.8}$Se$_2$, indicating replacement of Fe in the FeSe layer, presented in Table \ref{table}. Obtained small single crystals of these precursors are used to exchange K with LiOH hydrothermally as described in our earlier work\cite{Zhou_LiOH}. For such conversion,  it is crucial to include additional Fe powders and Sn metal plates in the autoclave to avoid formation of iron oxides and complete destruction of superconductivity. In this scheme, because the transition metal dopant, $M$, prefers covalent coordination with OH$^{-}$ group, they are driven from the FeSe layer of the K$_{0.85}$Fe$_{1.8-z}M_{z}$Se$_2$ to replace Li in the LiOH layer\cite{Sun_LiOH, Woodruff}. This is because the hard Lewis acid, Mn$^{2+}$, prefers coordination with the hard Lewis base of OH$^{-}$ as opposed to the soft base S$^{2-}$; this prevents the formation of impurity manganese sulfides. The vacancies in the FeSe layer are then supplemented by Fe$^{2+}$ from the solution. After the hydrothermal conversion, we obtain the products in both powder and single crystal forms.

After synthesis, we needed to determine the precise composition and crystallographic occupancy of all dopants. There has been multiple works to date on similar transition metal doping in the (LiOH)FeSe system, but none have been able to quantify the crystallographic location of the transition metal dopant \cite{gu2018sign, du2018sign, zhou2017doping, mao2018effect}. The location of the transition metal dopant is incredibly important in the realization of different physical properties as superconductivity in the iron chalcogenide systems is very sensitive to doping \cite{McQueen, Sun_LiOH, Woodruff}. To identify the crystallographic location of the Mn doping, we use extensive x-ray and neutron diffraction. Since, X-ray form factors of Mn are very close to Fe, it is extremely difficult to obtain their site occupancy using regular X-ray diffraction whereas elemental analysis methods are not able to tell where Mn is located nor whether they are from impurities. Therefore, high-resolution synchrotron powder x-ray diffraction (PXRD) has been paired with neutron powder diffraction (NPD) to fully understand the composition of these materials. Fig. \ref{fig:xrd_npd} shows powder x-ray diffraction and powder neutron diffraction data for (Li$_{1-x-y}$Fe$_{x}$Mn$_{y}$OD)FeSe. High resolution synchrotron PXRD allows for high quality Rietveld refinements of crystallographic structural parameters, Fe vacancies in the FeSe layer and total transition metal (M$^{*}$ = Fe + Mn) occupancies in the LiOH layer; however it does not allow for discrimination of transition metal dopants on the same crystallographic site. Thus, NPD complements synchrotron PXRD for high contrast between Fe and other transition metals occupying the same site.

Figure \ref{fig:xrd_npd}a shows powder x-ray diffraction data and corresponding Rietveld fit for (Li$_{1-x-y}$Fe$_{x}$Mn$_{y}$OD)FeSe. Refinements with PXRD were limited to occupation of total transition metal (M$^{*}$) doping in the hydroxide layer as well as total transition metal in the FeSe layer yielding a formula: (Li$_{1-x}M^{*}_{x}$OD)$M_{y}$Se. Rietveld refinements of the powder x-ray diffraction yielded a composition of  (Li$_{0.875(2)}M^{*}_{0.125(2)}$OD)$M$Se with lattice parameters $a$ = 3.8008(1) \AA \text{} and $c$ = 9.2394(2) \AA \text{} which is in close agreement with previous works \cite{Lynn_LiOH, Sun_LiOH, Zhou_LiOH, Chen_LiOH}. 


These refined structural and occupation values were then used as the starting model for the NPD. As mentioned, NPD was pivotal in determining the different transition metal dopants in the hydroxide layer and FeSe layers. Figure \ref{fig:xrd_npd}b,c show the NPD and corresponding Rietveld analysis for (Li$_{0.875(2)}M_{0.125(2)}$OD)$M$Se at 2 and 50 K respectively. Subsequent Rietveld analysis yielded a composition of (Li$_{0.875(2)}$Fe$_{0.062(3)}$Mn$_{0.062(3)}$OD)FeSe. Interestingly, it was found that all the manganese dopant migrates to the hydroxide layer with no refineable amount of manganese in the FeSe layer. It is understood that dopants in the FeSe layer destroys superconductivity \cite{Zhou_LiOH, Sun_LiOH}, so this reaction scheme allows for the superconductivity to be marginally altered while the magnetic layer can be manipulated. 




\section{Magnetic and transport properties}

\begin{figure*}[t!]
	\includegraphics[width=1.75\columnwidth]{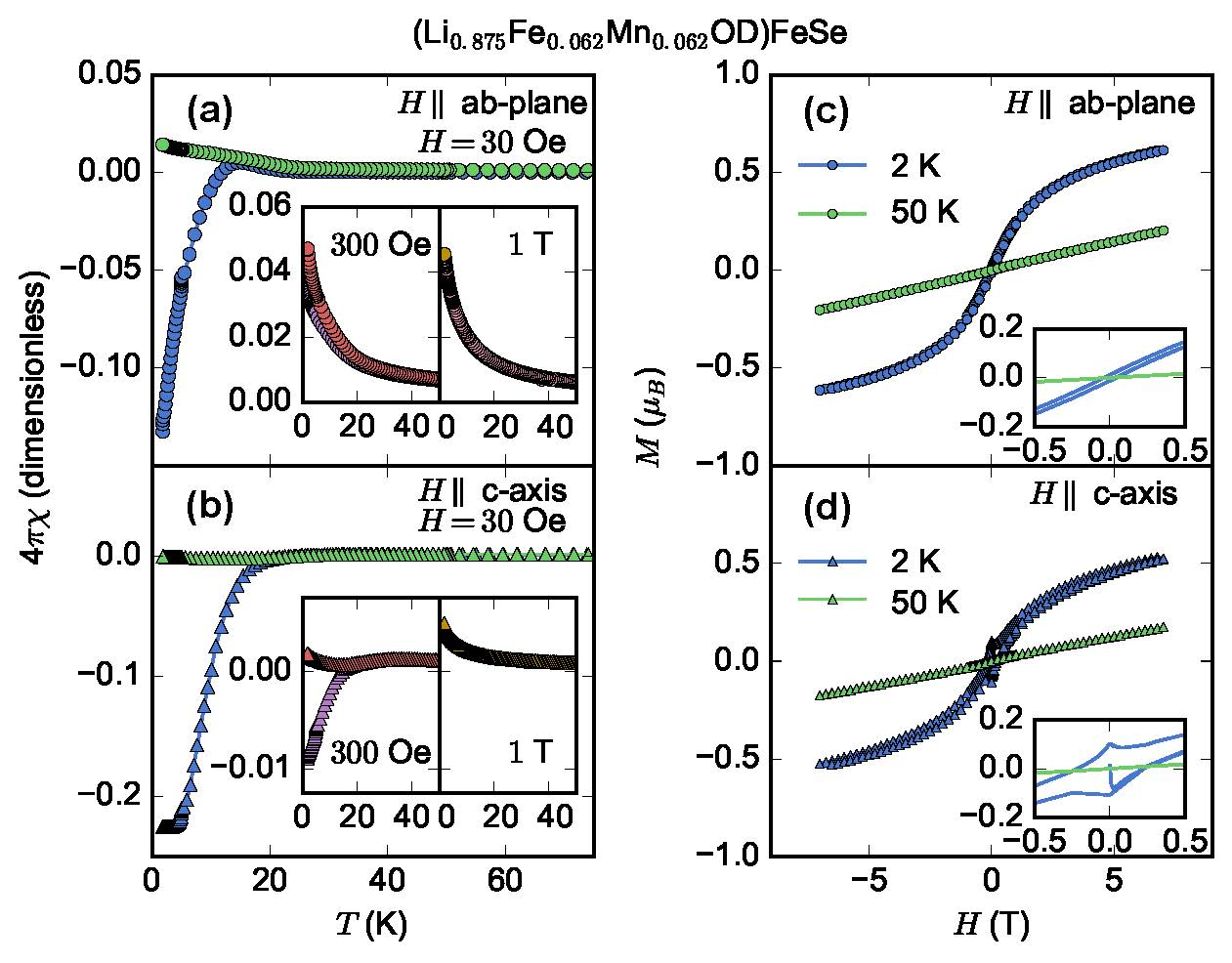}
	\caption{ Magnetic property measurements on an aligned single crystal of (Li$_{0.875(2)}$Fe$_{0.062(3)}$Mn$_{0.062(3)}$OD)FeSe. a)/b) Magnetic susceptibility data at various applied fields, aligned parallel to the crystallographic $ab$-plane and $c$-axis respectively,  showing a superconducting transition at 25 K but no second magnetic transition at lower temperature. c)/d) Isothermal magnetization at 2 K and 50 K, with the same orientation convention, showing magnetic hysteretic behavior superimposed on a superconducting signal at 2 K and paramagnetic behavior at 50 K. Note: (1 Oe = (1000/4$\pi$) A/m)}
	\label{fig:xvt}
\end{figure*}

\begin{figure}[t!]
	\includegraphics[width=0.95\columnwidth]{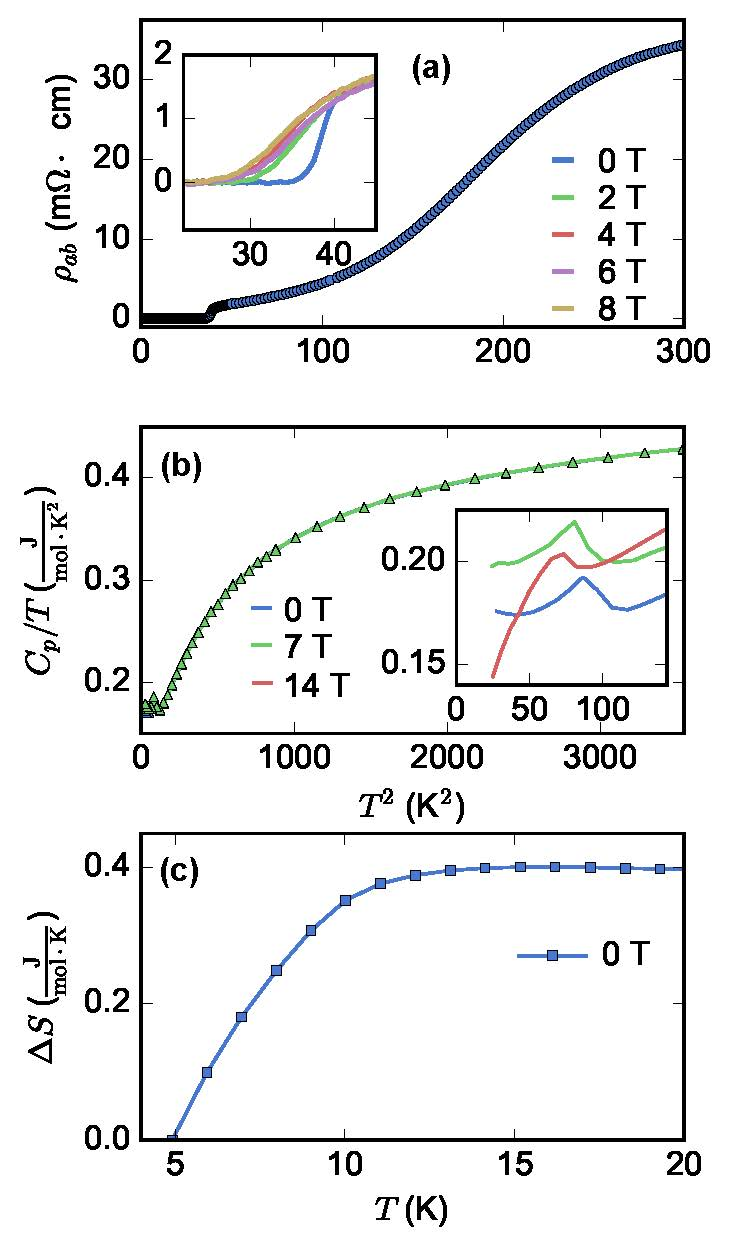}
 	\caption{ Electronic and thermal transport measurements on a single crystal of (Li$_{0.875(2)}$Fe$_{0.062(3)}$Mn$_{0.062(3)}$OD)FeSe. a) Temperature dependence of electrical resistivity with mutiple applied fields, with the applied field parallel to the crystallographic $c$-axis, showing a sharp superconducting transition at 41 K. At higher applied fields, the transition onset does not change, but the transition width is increased drastically. b) Temperature dependence of heat capacity at various applied fields showing no observable anomaly at the superconducting transition temperature, however; a magnetic transition is observed at $\sim$ 9 K that is weakly suppressed with high field (inset). c) The change in entropy calculated as described in main text.}
  	\label{fig:transport}
\end{figure}

 We have measured magnetic susceptibility at various applied fields and isothermal magnetization at 2 K and 50 K on a single crystal of  (Li$_{0.875(2)}$Fe$_{0.062(3)}$Mn$_{0.062(3)}$OD)FeSe. Figure \ref{fig:xvt}a/b show temperature dependent magnetic susceptibility, corrected for demagnetization factors of a two-dimensional plate \cite{prozorov}, with applied field applied parallel to the crystallographic $ab$-plane and $c$-axis respectively. Interestingly, the superconducting onset temperature is 25 K which is significantly lower than in the un-doped system. The highest observed shielding fraction after correction for demagnetization factors is $\sim$ 25\% which may indicate non-bulk superconductivity in this system. Within the (LiOH)FeSe system, many factors that affect the observable superconducting temperature: lattice parameters, transition metal concentration in the hydroxide and selenide layers, and structural homogeneity. \cite{Zhang_LiOH, Zhou_LiOH, Sun_LiOH}

Unlike previous reports, \cite{Lu_LiOH, Pachmayr} no magnetic transition is observed in the magnetic susceptibility even at higher applied fields. Interestingly, the only reports to show a transition in magnetic susceptibility below the superconducting transition are on powder samples via hydrothermal method \cite{Lu_LiOH, Pachmayr, Zhou_LiOH, Urban_LiOH}. Single crystals samples from the ion-exchange hydrothermal method fail to show the secondary magnetic transition below the observed superconducting transition \cite{Zhou_LiOH, khasanov, wang_LiOH, Du_LiOH, Zhang_LiOH,wang2_LiOH}. We hypothesize that sample uniformity in doping and structure caused by the different reaction conditions is likely the cause of these property differences. In the bottom-up synthesis, FeSe and LiOH layers are formed in-situ contrary to the ion-exchange method which requires the removal of interlayer ions before replacing with LiOH extended solid which would lead to more  structural and compositional variation.

We observe strong anisostropy in the magnetic susceptibility. Fields applied parallel to the crystallographic $ab$-plane have a much stronger effect on suppressing the superconducting transition than fields applied parallel to the $c$-axis as shown in the insets of Figure \ref{fig:xvt}a/b. This is consistent with previous reports \cite{Zhang_LiOH,  Dong_LiOH} and the observed transition is suppressed at relatively low fields, 300 Oe. The supression is due to the superconducting irreversibility temperature which is unique in these samples due to the very large vortex liquid region caused by extremely high anisotropy and two-dimensionality \cite{wang_LiOH}. 

We see an effect of underlying magnetic order is observable in isothermal magnetization measurements shown in Figure \ref{fig:xvt}c/d at 2 K and 50 K. For both field orientations, magnetization as a function of fields shows a superconducting signal superimposed on a weak magnetic background. Mirroring the behavior of magnetic susceptibility, the Meissner shielding is more clearly observable at low fields for the field applied parallel to the $c$-axis as compared to field applied parallel to the $ab$-plane. Again, this fast suppression of Meissner shielding could be due to true anisotropy, the filamentary nature of the superconductivity, or the effects of vortex dynamics in the sample. At higher temperatures, 50 K, the magnetic and superconducting signals are absent and paramagnetic behavior is observed. This paired with x-ray and neutron diffraction data shows that the magnetic signal is intrinsic to the system.

Electrical and thermal transport measurements shed more light on the superconducting and magnetic order in this system. Temperature dependence of electrical resistivity on a single crystal of  (Li$_{0.875(2)}$Fe$_{0.062(3)}$Mn$_{0.062(3)}$OD)FeSe, shown in Figure \ref{fig:transport}a, shows a sharp superconducting transition at 41 K in zero field. With field applied parallel to the crystallographic $c$-axis, it is observed that the superconducting transition onset is not changed, but the transition width in greatly increased at higher applied fields. This transition width broadening matches previous reports for this system. Specifically where this behavior has been justified by strong vortex flow as well as a wide temperature range for vortex liquid behavior due to anisotropy and two-dimensionality. \cite{wang_LiOH, wang2_LiOH, Zhang_LiOH, Dong_LiOH, Lin_LiOH, Wang3_LiOH, yi_vortex}

The superconducting onset temperature measured by resistivity is much higher than in the same crystal measured in magnetic susceptibility. This phenomena was observed in previous works as well and can be attributed to the vortex dynamics in the system \cite{Zhang_LiOH, wang2_LiOH, Du_LiOH, yi_vortex}. The residual resistivity ratio $\rho(300 \mathrm{ K})/ \rho(T_{c})$ for the sample presented is $\sim$ 25 which is higher than in previous reports \cite{Zhang_LiOH, wang2_LiOH, Lin_LiOH, Wang3_LiOH}. The observed normal state behavior is non-linear which has been demonstrated in previous works \cite{Zhang_LiOH, wang2_LiOH, Lin_LiOH, Wang3_LiOH} and has been attributed to over-doping in the sulfide analogue \cite{zhou}.

\par Due to the very high critical field in these samples \cite{Dong_LiOH}, heat capacity measurements were employed to evaluate the effect magnetic ordering has within the superconducting regime. Heat capacity measurements on this system fail to show anomalous behavior around the superconducting transition \cite{Lu_LiOH, Topping_LiOH, Lu_LiOH2}. Temperature dependence of heat capacity at various applied fields shown in Figure \ref{fig:transport}b shows similar behavior. No anomaly is observed at the superconducting transition temperature, however; a magnetic transition is observed at $\sim$ 9 K in zero applied field parallel to the crystallographic $c$-axis. In these previous reports, \cite{Lu_LiOH, Topping_LiOH, Lu_LiOH2} the magnetic transition was suppressed at high applied fields which was justified by antiferromagnetic or spin-glass ordering in the system. However, these measurements were done on samples made via the bottom-up hydrothermal synthesis, which behave differently than those produced via the ion-exchange method. To that end, the magnetic transition shown in Figure \ref{fig:transport}b shows slight temperature suppression at 14 T, but is still clearly observable. We can conclude that the magnetic order in this system is of an antiferromagnetic or ferrimagnetic nature when combined with the observation of satellite reflection in neutron diffraction.

To calculate the magnetic contribution to the heat capacity at low temperatures, the temperature dependent $C_{p}/T$ data at zero applied field was fit to a third order polynomial above 20 K. This fitting was then extrapolated to low temperatures and subsequently subtracted from the raw data below 20 K. The remainder was then integrated to yield $\Delta S$ as shown in Figure \ref{fig:transport}c as a function of temperature. The change in entropy sharply increases up to 12 K before leveling off at a value of 0.38(8) J/(mol$\cdot$K), which is significantly lower than the expectation for a free electron spin. The entropy change through a magnetic transition is defined as $\Delta S$ = $cR\mathrm{ln}(2J + 1)$, where $c$ is the concentration of magnetic ions in the system and $R$ is the ideal gas constant. Our compositional analysis of single crystals of (Li$_{0.875(2)}$Fe$_{0.062(3)}$Mn$_{0.062(3)}$OD)FeSe showed a concentration of magnetic ions in the hydroxide layer as $c$ = 0.125, yielding an effective $J$ = 0.22(6) for the system and effective moment $\mu_{\mathrm{eff}}$ = 1.05(2) $\mu_{\mathrm{B}}$, spin-only. 

The expected spin values for Fe$^{2+}$ and Mn$^{2+}$ in tetrahedral coordination (quenched orbital angular momentum) are 2 and 5/2 respectively. These values are much larger than the observed 0.22(6) from entropy measurements. Previous reports observed similar behavior and concluded that this spin underestimation was due to spin-glassy character of the magnetism in the hydroxide layer \cite{Topping_LiOH}. The robustness of the observed transition up to 14 T trends away from a spin-glass description for the magnetic behavior but spin glassiness cannot be ruled out \cite{Lu_LiOH, Topping_LiOH, binder}. In-depth analysis at the atomic level through imaging or spectroscopy would be useful to understand how the dilute magnetic ions in the hydroxide layer order to from long range correlations.


\section{Conclusions}

The coexistence of superconductivity and magnetism may be the key to realize the next generation of multi-functional materials. To that end, we have successfully synthesized a series of late transition metal doped (LiOH)FeSe single crystals through a hydrothermal ion-exchange reaction has been shown to house the co-existence of superconductivity and long-range magnetic order in the case of the nominally 20\% Mn-doped sample. We have fully characterized the 20\% Mn-doped sample through powder x-ray and neutron diffraction yielding a stoichiometry of (Li$_{0.875(2)}$Fe$_{0.062(3)}$Mn$_{0.062(3)}$OD)FeSe showing that the doped Mn migrates fully to the hydroxide layer. Magnetic measurements show a superconducting transition at 25 K with shielding fraction $\sim 25$\% indicating the superconductivity in single crystal samples may be filamentary in nature with magnetization showing clear Meissner shielding and magnetic signal co-existing at low temperatures exclusively. Transport measurements show a sharp 41 K transition in resistivity with a distinct non-linear normal state with no clear anomaly detected around the superconducting transition in heat capacity. However, a clear magnetic peak was observed at $\sim 9$K in heat capacity measurement which remains robust up to 14 T applied field parallel to the crystallographic $c$-axis. Powder neutron diffraction data reveals a satellite magnetic reflection at $Q = 1.41 - 1.44 \text{ }\mathrm{\AA}^{-1}$ indicating the long-range character of the magnetic order in this system observed for the first time. Additional neutron experiments would be required to elucidate the nature of the magnetic ordering in this system. This work shows that the (LiOH)FeSe system offers a platform for chemical manipulation to induce the coexistence of long range magnetic order and superconductivity for possible exploitation as multi-functional materials and for use in quantum computing.

\newpage
\bibliographystyle{apsrev4-1}

%

\section{Acknowledgements}
Research at the University of Maryland was supported by the NSF Career DMR-1455118, the AFOSR Grant No. FA9550-14-10332, and the Gordon and Betty Moore Foundation Grant
No. GBMF4419. We also acknowledge support from the Center for Nanophysics and Advanced Materials. We acknowledge the support of the National Institute of Standards and Technology, U.S. Department of Commerce, in providing the neutron research facilities used in this work and the NIST awards 70NANB12H238 and 70NANB15H261. The identification of any commercial
product or trade name does not imply endorsement or recommendation by the National Institute of Standards and Technology. The part of the research that was conducted at ORNL's
High Flux Isotope Reactor was sponsored by the Scientific User Facilities Division, Office of Basic Energy Sciences, US Department of Energy. We thank S.A.J. Kimber for his assisstance with HB-2A measrements. The use of the Advanced Photon Source at Argonne National Laboratory was supported by the U. S. Department of Energy, Office of Science, Office of Basic Energy Sciences, under Contract No. DE-AC02-06CH11357. We thank S. Lapidus for his assistance with the 11-BM measurements.

\pagebreak
\widetext
\begin{center}
\textbf{\large Supplemental Materials: Long range magnetic order in hydroxide layer doped (Li$_{1-x-y}$Fe$_{x}$Mn$_{y}$OD)FeSe}
\end{center}
\setcounter{equation}{0}
\setcounter{figure}{0}
\setcounter{table}{0}
\setcounter{page}{1}
\makeatletter
\renewcommand{\theequation}{S\arabic{equation}}
\renewcommand{\thefigure}{S\arabic{figure}}
\renewcommand{\thetable}{S\arabic{table}}

\begin{figure*}[thb!]
	\includegraphics[width=4in, height=5in]{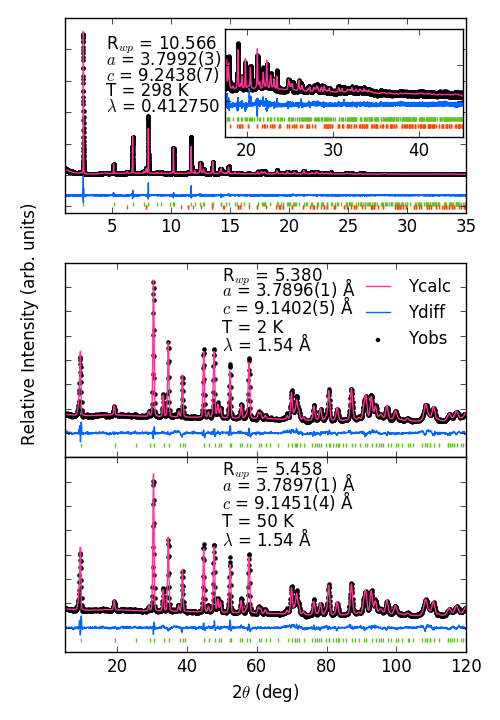}
 	\caption{ (top) High resolution synchrotron powder X-ray diffraction patterns for (Li$_{1-x-y}$Fe$_{x}$Mn$_{y}$OD)Fe$_{1-z}$Mn$_{z}$Se collected at room temperature, (middle and bottom) show powder neutron diffraction data for (Li$_{1-x-y}$Fe$_{x}$Mn$_{y}$OD)Fe$_{1-z}$Mn$_{z}$Se at 2 K and 50 K respectively. Green tick marks represent the targeted tetragonal layered phase and orange tick marks represent impurity selenium. These are shown below the calculated, observed, and differences curves from Rietveld analysis. Analysis yielded a composition of (Li$_{0.888(3)}$Fe$_{0.072(4)}$Mn$_{0.038(2)}$OD)Fe$_{0.980(5)}$Se.}
  	\label{fig:neutron_10Mn}
\end{figure*}

\begin{table*}[thb!]
\centering
\begin{tabular}{|l|l|l|l|l|}
\hline
                 & \multicolumn{2}{l|}{10\% Nominal Doping}                  & \multicolumn{2}{l|}{20\% Nominal Doping}                  \\ \hline
Transition Metal & a lattice parameter (\AA) & c lattice parameter (\AA) & a lattice parameter (\AA) & c lattice parameter (\AA) \\ \hline
Mn               & 3.8963                      & 14.1434                     & 3.8937                      & 14.1513                     \\ \hline
\end{tabular}
\caption{Refined lattice parameters of K$_{0.85}$Fe$_{1.8-x}$Mn$_{x}$Se$_{2}$  starting materials made via a high temperature self flux reactions from the corresponding elements. The 10\% and 20\% nominal doping come from the stoichiometric additons of each element to the growth. The reported lattice parameters for the Fe (un-doped) version come from a similarly targeted stoichiometric compound of K$_{0.85}$Fe$_{1.9}$Se$_{2}$ via Shoemaker, D.P., \textit{et al}.  Phys. Rev. B 86.18 (2012): 184511.}
\label{table}
\end{table*}

\end{document}